\documentclass[iop]{emulateapj}
\newcommand {\Lya}    {Ly$\alpha$}   

\newcommand {\HI}     {\ion{H}{1}}   
\newcommand {\OVI}    {\ion{O}{6}}   

\newcommand {\CIV}    {\ion{C}{4}}
\newcommand {\SiIV}   {\ion{Si}{4}}
\newcommand {\SiIII}  {\ion{Si}{3}}
\newcommand {\SiII}   {\ion{Si}{2}}
\newcommand {\SII}    {\ion{S}{2}}

\newcommand {\FeII}   {\ion{Fe}{2}}
\newcommand {\AlII}   {\ion{Al}{2}}
\newcommand {\CII}    {\ion{C}{2}}
\newcommand {\CaII}   {\ion{Ca}{2}}
\newcommand {\NaI}    {\ion{Na}{1}}

\newcommand {\kms}    {km~s$^{-1}$}
\newcommand {\NHI}    {$N_{\rm HI}$}

\newcommand {\tnma}{\tablenotemark{a}}
\newcommand {\tnmb}{\tablenotemark{b}}
\newcommand {\tnmc}{\tablenotemark{c}}

\newcommand {\etal}  {et~al.}

\slugcomment{Accepted for publication in  the Astrophysical Journal; May 28, 2015}
\usepackage{color}
\begin{document}
\shorttitle{Gas in the M\,31 Stellar Stream}
\shortauthors{A. Koch et al.}
\title{A Detection of Gas Associated with the M\,31 Stellar Stream\altaffilmark{$\dagger$}}
\author{
Andreas Koch\altaffilmark{1},
Charles W. Danforth\altaffilmark{2}, 
R. Michael Rich\altaffilmark{3},  
Rodrigo Ibata\altaffilmark{4}, \and
Brian A. Keeney\altaffilmark{2}
}
\altaffiltext{$\dagger$}{Based on observations made with the NASA/ESA {\it Hubble Space Telescope}, obtained from the data archive at the Space Telescope Science Institute. STScI is 
operated by the Association of Universities for Research in Astronomy, Inc. under NASA contract NAS 5-26555.}
\altaffiltext{1}{Zentrum f\"ur Astronomie der Universit\"at Heidelberg,  Landessternwarte, K\"onigstuhl 12, 69117 Heidelberg, Germany}
\altaffiltext{2}{CASA, Department of Astrophysical and Planetary Sciences, University of Colorado, 389-UCB, Boulder, CO 80309}
\altaffiltext{3}{Physics and Astronomy Building, 430 Portola Plaza, Box 951547, Department of Physics and Astronomy, University of California, Los Angeles, CA 90095}
\altaffiltext{4}{Observatoire de Strasbourg, 11, rue de l'Universit\'e, F-67000, Strasbourg, France}
%
%
%
\begin{abstract}
Detailed studies of stellar populations in the halos of the Milky Way and the Andromeda (M\,31) galaxies have shown increasing numbers of tidal streams and dwarf galaxies,
 attesting to a complicated and on-going process of hierarchical structure formation.  The most prominent feature in the halo of M\,31 is the Giant Stellar Stream, a structure $\sim$$4.5$\degr\ in extent along the sky, 
 which is close to, but not coincident with the galaxy's minor axis.  The stars that make up this stream are kinematically and chemically distinct from the other stars in the halo.  
Here, we present HST/COS high-resolution ultraviolet absorption spectra of three Active Galactic Nuclei 
 sight lines which probe the M\,31 halo, including one that samples gas in the main southwestern portion of the Giant Stream.  We see two clear absorption components in many metal species at velocities typical of the M\,31 halo and a third, blue-shifted component which arises in the stream.  Photoionization modeling of the column density ratios in the different components shows gas in an ionization state typical of that seen in other galaxy halo environments { and suggests solar to slightly super-solar metallicity, 
consistent with previous findings from stellar spectroscopy. } 
\end{abstract}
\section{Introduction}
The assembly of galaxies via hierarchical accretion -- wherein smaller structures merge to create larger structures -- is now a well-established doctrine of galaxy formation, which only requires the action of gravity. 
This universal prediction of cosmological simulations \citep[e.g.,][]{Boylan-Kolchin10} is also well observed in the local universe, with simulations predicting a multitude of substructures in the Milky Way (MW) halo \citep{HelmiWhite99}.  
Streams of stars and gas have been observed in the MW and have been linked to previous accretion of satellite galaxies \citep{SearleZinn78,Ibata94,Belokurov07}.  The Magellanic Stream \citep{Mathewson74} observed in 21-cm emission is the result of the interaction between the Large and Small Magellanic Clouds (LMC and SMC, respectively) and the MW and may, gigayears hence, result in the merger of the Magellanic and MW galaxies.  On the stellar side, the advent of the Sloan Digital Sky Survey \citep{York00} has led to an explosion in the discovery of stellar streams in the MW halo \citep[e.g.,][]{Yanny03,Belokurov07,Grillmair09,Newberg09,Martin13}, all of which presumably have their origin in minor mergers in the Galaxy's past.

In the past decade our knowledge of the structure of our nearest large galactic neighbor, M\,31, has undergone a similar transformation.  Deep, wide-field imaging by \citet{Ibata01} revealed a large stream of stars extending from near the center of M\,31 several degrees to the south east close to the minor axis of the galaxy.  Subsequent studies \citep{Ferguson02,McConnachie03} found that this Giant Stellar Stream (GSS) exists as a coherent structure 
more than 4\degr\ on the sky, inclined at $\sim$60\degr~with respect to the line of sight,  
and extending over $\ga140$ kpc.  The GSS is aligned with the major M\,31 satellites M\,32 and NGC\,205 and, while it is tempting to associate the stream with those companions, modeling shows that it most likely represents the { trailing arm left} from a recent, radial
 merger of a galaxy of $3-5\times 10^9$ M$_\odot$ (Mori \& Rich 2008, Fardal et al. 2013).   Alternatively it may be the fossil remnant 
of a cannibalized satellite or a tidal bridge connecting to its neighboring spiral M\,33. 

 \citet{Brown06} found from Hubble Space Telescope (HST)  color-magnitude diagrams that the Stream and halo fields are indistinguishable, which suggests that only one progenitor was responsible for the debris deposited in inner halo regions.  However, if dynamical mixing were efficient, it could erase the signatures from different infall sources.  Further deep imaging and stellar spectroscopic studies have shown that the halo of the M\,31-M\,33 system is far more complicated than originally thought and hints at an eventful formation history \citep[e.g.,][]{Ibata07,McConnachie09}.  In addition to the GSS, many other structures have been discovered within the halo of M\,31.  Some of these structures are associated with the large and growing population of Andromeda satellite galaxies \citep{Richardson11} and globular clusters \citep{Mackey10}, while others are less clearly connected.
{ It has been suggested that many of the stream-like features in the halo could result from the heating of M31's thin disk that has been ejected during a recent merger (e.g., Richardson et al. 2008; Bernard et al. 2015).}

We know relatively little about the progenitor of the GSS.  Modeling of the encounter leads to mass estimates ranging from $3\times 10^8$ M$_\odot$ (Hammer et al. 2010) to $<5\times 10^9$ M$_\odot$ (Mori \& Rich 2008); the most recent estimate of $3\times 10^9$ M$_\odot$  comes from Fardal et al. (2013), placing the system at the mass of the LMC.  
{ It has been suggested that the lack of a compact remnant favors a disk-like progenitor. 
Furthermore, Fardal et al. (2008) argue that asymmetries in the surface brightness distribution across the GSS and its morphology in terms 
of a more compact core with an extended envelope can best be explained if this progenitor disk galaxy  was rotating.
Based on the metal-rich nature and the age-metallicity relation of several halo fields with stream-like color magnitude diagrams Bernard et al. (2015) 
suggest an early-type galaxy as a candidate, which experienced rapid chemical enrichment to Solar abundances already by $z=1$.

However, no neutral gas analogous to the Magellanic Stream  is evident in the immediate vicinity of the GSS (Braun et al. 2003). 
A recent re-analysis of the H\,{\sc i} data by Lewis et al. (2013) showed the presence of gas filaments, bridging M\,31 and  the near-by spiral M\,33, 
although they are significantly offset from the stellar stream. 
Lewis et al. (2013)  also identified a number of individual gas clouds, but noted a remarkable lack of any spatial correlation between  stellar and gaseous substructures, 
arguing for several dynamical processes at play that affect both components in different ways. The only gas associated with the GSS thus appears at the point where it passes the disk of M\,31.
}

Absorption-line spectroscopy provides a different and 
{ complementary}
 tool for analyzing the structure and kinematics of circumgalactic 
 { matter (e.g., Rao et al. 2013)}.
 Stellar spectroscopy provides a representative view of stellar kinematics and overall metallicities, while absorption spectroscopy of background sources gives kinematic and chemical information about the {\em gas phase} in the system.  
In this paper, we present far-ultraviolet (FUV) spectra of three active galactic nuclei (AGN) obtained with the {Cosmic Origins Spectrograph} ({COS}) onboard the HST.  These three sight lines probe the M\,31 halo at projected distances of 26--51 kpc in three different 
quadrants of the system (Fig.~1).  
\begin{figure}
\includegraphics[width=1.1\hsize,angle=270]{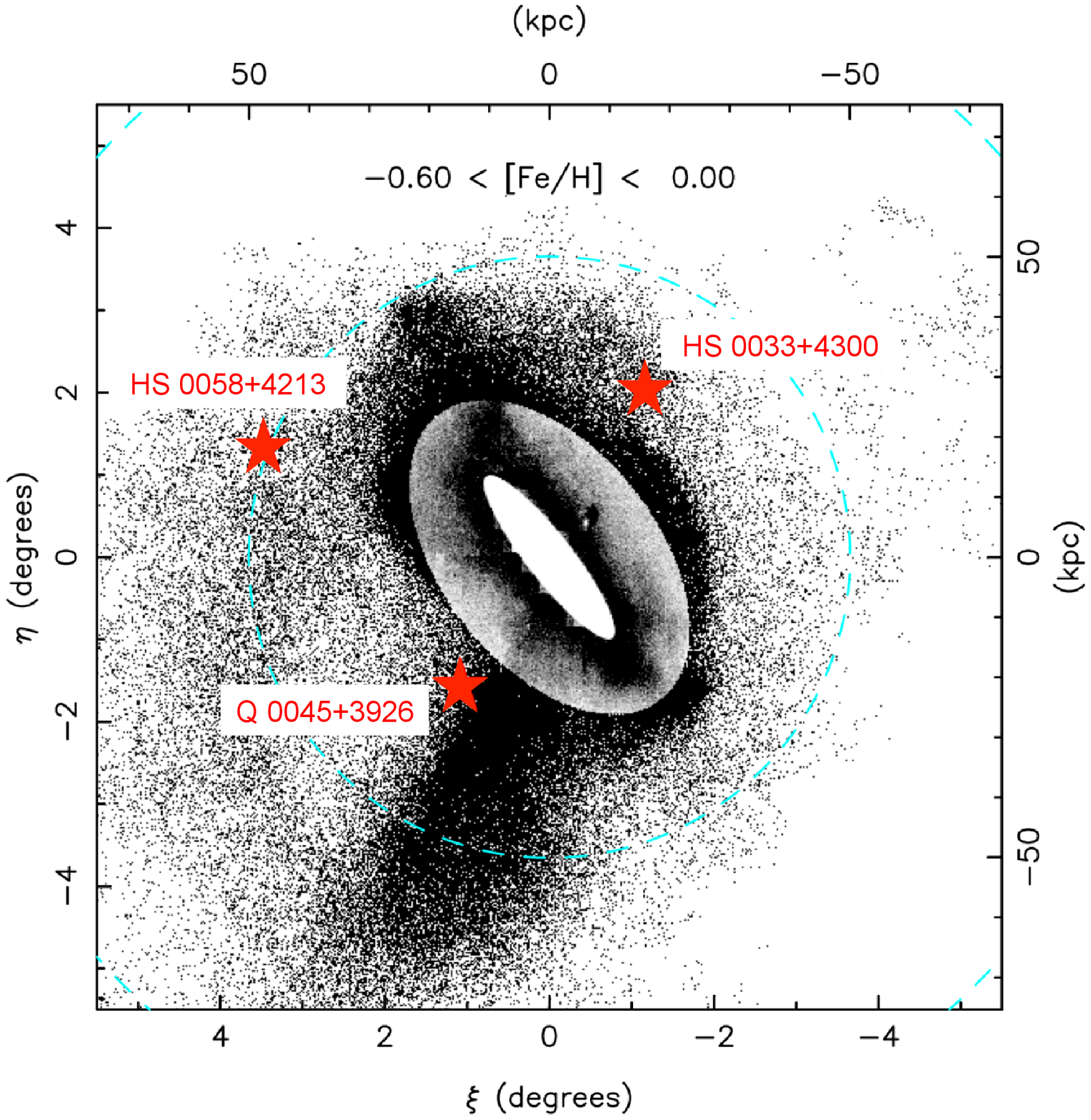} 
\caption{The locations of the three AGN used in this study overlaid
{ on a density map of metal-rich stars (after Ibata et al. 2014).}
Notice that Q\,0045+3926 lies near the edge of the Giant Stellar Stream.
Here, we have masked out the inner disk of M\,31 within a radius of 17 kpc. 
The inner halo out to a radius of 30 kpc and at an  axis ratio of 3:5 is displayed at lower contrast to render visible the inner halo substructure.
}
\end{figure}

We see strong absorption from gas near the systemic velocity of M\,31 in all three sight lines.  The sight line that probes the GSS additionally shows clear absorption at more negative velocities, near the systemic velocity of the stellar members of the Giant Stream. 
While Lehner et al. (2015) associate such  absorbers with the hot, gaseous corona of M\,31, no connection to the stellar component of the GSS has been made. 

Our measurements of the M\,31 and GSS-velocity gas are presented in \S 3.  In \S 4, we compare the observed gas-phase detections with the stellar kinematics in the same field and 
a kinematic model from the literature. { Here, we also  analyze the absorbers by application of basic photoionization models.} 
 Our conclusions are presented in \S 5.
\section{Observations and Data Reduction}
We base our analysis on HST/COS \citep{Green12,Osterman11} observations of three AGN sight lines, which pass near M\,31 (Fig.~1).  Of the three sight lines, the one near the GSS (Q\,0045+3926) is noteworthy as it lies 
 0.2\degr  { ($\sim$2.7 kpc)} East of the main stellar body of the GSS.  The predominance of stellar stream features and complexity in all regions of the halo is well documented (e.g. McConnachie et al. 2009).  
Three Seyfert galaxies were observed in 2009, 6--7 October as part of Program 11632 (P.I. R.M. Rich).  Each observation consisted of a single exposure in the COS/G130M ($1135<\lambda<1450$ \AA;  6.00 ksec) grating and one to three exposures with COS/G160M ($1400<\lambda<1795$ \AA).  In addition, Q\,0045$+$3926 was observed in late 2010 (P.I. N. Arav).  
Observation details are given in Table~1.  
\begin{deluxetable*}{ccccccccc}[htb!]
\tabletypesize{\footnotesize}
\tablecolumns{9} 
\tablewidth{0pt} 
\tablecaption{Summary of COS Observations}
\tablehead{
  \colhead{Target Name} & 
  \colhead{Alternate Name} &
  \colhead{$\alpha$} &
  \colhead{$\delta$} &
  \colhead{$z_{\rm AGN}$} & 
  \colhead{$\rho_{\rm M\,31}$} &
  \colhead{Date of} &
  \colhead{Spectral} & 
  \colhead{$t_{\rm exp}$} \\
  \colhead{ } &
  \colhead{ } &
  \colhead{ (J2000.0)} & \colhead{ (J2000.0)} &
  \colhead{ } &
  \colhead{[kpc]\tnma} &
  \colhead{of observation} &
  \colhead{Grating} & 
  \colhead{[s]}
}
\startdata
Q\,0045$+$3926  & IO\,And      & 00:48:19.0 & $+$39:41:12 & 0.134 &  26.1 &  2009/10/06 & G130M\tnmb  &  2330 \\ 
                &              &            &	          &	  &       &  	        & G160M\tnmb  &  2900 \\ 
                &              &            &	          &  	  &       &  2010/09/07 & G160M\tnmc  & 14259 \\ 
                &              &            &	          &	  &       &  2010/10/19 & G130M\tnmc  & 11155 \\ 
HS\,0033$+$4300 & \nodata      & 00:36:23.0 & $+$43:16:40 & 0.120 & 31.8  &  2009/10/06 & G130M\tnmb  &  2455 \\ 
                &              &            &             &       &       &             & G160M\tnmb  &  9145 \\ 
HS\,0058$+$4213 &J\,0101$+$4229& 01:01:31.2 & $+$42:29:35 & 0.190 & 50.7  &  2009/10/07 & G130M\tnmb  &  2320 \\ 
                &              &            &             &       &       &             & G160M\tnmb  &  2970 
\enddata
\tablenotetext{a}{Minimum impact parameter from sight line to center of M\,31 based on the scale 228 pc arcmin$^{-1}$ and mean distance $792\pm193$ kpc.}
\tablenotetext{b}{Program 11632, P.I. Rich}
\tablenotetext{c}{Program 11686, P.I. Arav}
\end{deluxetable*}

The exposures were obtained from the Mikulski Archive for Space Telescopes.  The calibrated, one-dimensional spectra for each target were next coadded with the custom IDL procedures described in detail by \citet{Danforth10}.  The coaddition routines automatically scales exposures taken during different epochs to take into account source variability and stitch G130M and G160M data together into a continuous spectrum over the range of 1150--1750~\AA.  
{ Since the two objects HS\,0033$+$4300 and HS\,0058$+$4213 were observed with a single COS grating position, 
each dataset features a pair of $\sim20$ \AA\ gaps in the spectral coverage, corresponding to the gap between the two far-UV detectors.}

We fit continua to each of the data sets using a semi-automated line identification and spline-fitting technique as follows.  First, the spectra are split into 5--10 \AA\ segments.  Continuum pixels within each segment are identified as those for which the 
signal-to-noise (S/N) ratio 
vector is less than 1.5 $\sigma$ below the median S/N value for all the pixels in the segment.  Thus, absorption lines (flux lower than the segment average) are excluded, as are regions of increased noise (error higher than segment average).  The process is iterated to convergence and a spline is fitted through the mean continuum pixel value in each segment.  We check the continuum fit of each entire spectrum manually and the continuum region identifications are adjusted as needed.  The continuum identification and spline-fitting processes work reasonably well for smoothly varying data, but they were augmented with piecewise-continuous Legendre polynomial fits in a few cases.  In particular, spline fits perform poorly in regions of sharp spectral curvature, such as the Galactic \Lya\ absorption and at the peaks of cuspy emission lines.  Complete details on the process are given in Danforth et al. (2014).  

The region around M\,31 has previously been mapped in the 21\,cm emission line of \HI\ by several authors.  \citet{Thilker04} conducted a Green Bank Telescope (GBT) survey of the region, while \citet{BraunThilker04} present a deep survey in the M\,31 region made with the Westerbork Synthesis Radio Telescope (WSRT) operated in total-power mode \citep[see][for details]{Braun03}.  While the absorption-line data are much more sensitive than emission, the COS data give no information on the neutral hydrogen column density since the MW damped \Lya\ absorption profile extends over thousands of \kms.

Unfortunately, the velocity calibration of the UV spectra is uncertain 
to a much larger degree than radio emission data. HST spectra are 
recorded in a heliocentric velocity frame with an on-board wavelength 
calibration lamp. However, the precise wavelength solution varies by 
$\sim$0.2 \AA~($\sim$2 resolution elements) over the detector and the zero point is 
uncertain by $\sim$15 km\,s$^{-1}$ due to inaccuracies in centering the target in the 
COS aperture inherent to the target acquisition procedure. The coalition 
process cross-correlates strong absorption lines in each detector segment 
in each exposure with each other before coaddition, so the relative 
wavelength solution over a few \AA ngstroms is reasonably accurate. 

However, no cross-segment wavelength correlation is attempted.  We correct for these effects to first order by aligning narrow low-ionization metal absorption lines in the data in velocity space to the \HI\ 21-cm profiles from the 
{ Leiden-Argentine-Bonn survey \citep[]{LABsurvey}}. 
The \HI\ emission toward all three sight lines shows a peak at $0<v_{\rm LSR}<+3$ \kms\ with significant, distributed emission toward negative velocities.  
Fits to the Q\,0045$+$3926 
COS data show the reddest absorption component in several ions at $v\sim-20$ \kms.  Poor data quality in the other two spectra hampers this analysis rather severely, 
{ but the strongest Galactic absorption}
 appears to be at $-40\la v
\la-30$ \kms.  These values do not  vary considerably with rest wavelength, so we adopt a uniform velocity shift of $+20$ \kms\ for Q\,0045$+$3926 and $+35$ \kms\ for HS\,0033$+$4300 and HS\,0058$+$4213 to bring the data 
approximately into the Local Standard of Rest (LSR) reference frame.  Overall, we estimate a systematic velocity uncertainty of $\pm15$ \kms\ inherent in the absolute velocity of any absorption measurements henceforth. 

We measure absorption features in three ways.  { Firstly, } unsaturated absorption features with clear component structure can be fitted via the apparent optical depth (AOD) and profile-weighted mean quantity methods \citep{SavageSembach91,SembachSavage92}.  
{ Secondly, } multi-component Voigt profile fits often provide more accurate measurements in cases of mildly saturated lines.  The significance of any individual line fit is calculated based on the local S/N ratio, the Doppler $b$-parameter, and fitted equivalent width via the algorithms of \citet{Keeney12}.  Finally, we make use of {\em simultaneous} multi-component Voigt profile fits to multiple transitions of the same species, e.g., \SiII\ 1260.42, 1193.29, 1190.42, and 1526.70 \AA.  
{ These techniques produce equivalent results for absorption profiles of moderate strength, but each has its advantages in different circumstances:  Voigt fitting, which is generally preferred, 
can accurately measure lines which are moderately saturated ($\tau\sim$2--3).  The down-side of this technique is that a velocity component structure must be assumed.  
AOD measurements are well-suited for shallow, noisy absorbers or where the component structure is uncertain, while  
this method tends to under-estimate column densities for very strong absorbers. The latter does not pose a concern in our only mildly saturated absorbers.
For details on the consistencies between the methods, we refer the reader to Danforth \& Shull (2008; and references therein).} 
Measurements from all three techniques are performed on the data and the resulting values are presented in Tables~2 and 3, 
{ where we label the absorber ``components`` by the error-weighted mean of all species.}
\begin{deluxetable*}{lcccl}[htb!]
 \tabletypesize{\footnotesize}
 \tablecolumns{5} 
 \tablewidth{0pt} 
 \tablecaption{Species Measurements Toward Q\,0045$+$3926}
 \tablehead{
   \colhead{} &
   \colhead{Velocity $v$\tnma} &
   \colhead{$b$-value} &
   \colhead{$\log\,N$} &
   \colhead{} \\
   \raisebox{1.5ex}[-1.5ex]{Species} &
   \colhead{{ [km\,s$^{-1}$]} } &
   \colhead{{ [km\,s$^{-1}$]} } &
   \colhead{{ [cm$^{-2}$]} } &
   \raisebox{1.5ex}[-1.5ex]{Notes} 
   }
 \startdata
 \cutinhead{$v\approx-170$ \kms\ component}
  O\,I    &$ -170\pm5 $&$ 20:        $&$ 12.7:$\tnmb   & $<1\sigma$, AOD, Voigt  \\
  Si\,II  &$ -158\pm4 $&$ 21\pm7 $&$ 13.39\pm0.06 $& $\lambda 1190, 1193$; AOD, Voigt \\
  Si\,III &$ -162\pm2 $&$ 19\pm4     $&$ 13.17\pm0.11 $& AOD, Voigt \\
  Si\,IV  &$ -160\pm7 $&$ 22\pm5 $&$ 12.92\pm0.12 $& doublet; AOD, Voigt \\
  C\,II   &$ -130:    $&$ 30:        $&$ 14:      $& Voigt only, strongly blended \\
  C\,IV   &$ -173\pm2 $&$ 18\pm3 $&$ 13.57\pm0.03 $& AOD (doublet), Voigt ($\lambda 1548$ only) \\
  N\,V    &\nodata     &\nodata   &$<12.97        $& doublet; AOD upper limit  \\ 
 \cutinhead{$v\approx-230$ \kms\ component}
  O\,I    &$ -236\pm7 $&$ 20:        $&$ 13.1:$\tnmb   & AOD, Voigt ($\sim 1 \sigma$) \\
  Si\,II  &$ -222\pm16$&$ 24\pm7 $&$ 12.99\pm0.29 $& $\lambda 1190, 1193$; AOD, Voigt \\
  Si\,III &$ -228\pm4 $&$ 30\pm5     $&$ 13.16\pm0.04 $& AOD, Voigt \\
  Si\,IV  &$ -225\pm4 $&$ 28\pm4 $&$ 13.22\pm0.10 $& doublet; AOD, Voigt \\
  C\,II   &$ -227\pm3 $&$ 24\pm5     $&$ 13.89\pm0.02 $& AOD, Voigt \\
  C\,IV   &$ -230\pm3 $&$ 25\pm7 $&$ 13.98\pm0.07 $& doublet; AOD, Voigt \\
  N\,V    &$ -230\pm5 $&$ 10:    $&$ 12.97\pm0.12 $& $\sim3\sigma$, doublet; AOD, Voigt \\
 \cutinhead{$v\approx-370$ \kms\ component}
  O\,I    &$ -374\pm7 $&$ 20:        $&$ 13.4:$\tnmb  & AOD, Voigt ($\sim2\sigma$) \\
  Si\,II  &$ -359\pm4 $&$ 24\pm2 $&$ 12.88\pm0.10 $& $\lambda 1190, 1193, 1260$; AOD, Voigt \\
  Si\,III &$ -355\pm2 $&$ 35\pm6     $&$ 13.09\pm0.04 $& AOD, Voigt \\
  Si\,IV  &$ -390\pm9 $&$ 40:        $&$ 12.62\pm0.11 $& $\lambda 1393$ only; AOD, Voigt \\
  C\,II   &$ -351\pm5 $&$ 29\pm6     $&$ 14.12\pm0.05 $& AOD, Voigt \\
  C\,IV   &$ -378\pm1 $&$ 34\pm5 $&$ 13.20\pm0.01 $& AOD (doublet), Voigt ($\lambda 1548$ only) \\
  N\,V    &$ -393\pm16$&$ 30:    $&$ 12.65\pm0.06 $& $<3\sigma$; $\lambda 1238$ only; AOD, Voigt
 \enddata
 \tablenotetext{a}{All line profiles have been shifted by $+20$ \kms\ as discussed in the text.}
 \tablenotetext{b}{The $3\sigma$ detection limit for O\,I lines is $\log\,N<13.54$.  Marginal, lower-significance measurements are presented for completeness.}
\end{deluxetable*}
\begin{deluxetable*}{lcccl}[htb!]
 \tabletypesize{\footnotesize}
 \tablecolumns{5} 
 \tablewidth{0pt} 
 \tablecaption{Species Measurements Toward HS\,0033$+$4300 and HS\,0058$+$4213}
 \tablehead{
   \colhead{} &
   \colhead{Velocity $v$\tnma} &
   \colhead{$b$-value} &
   \colhead{$\log\,N$} &
   \colhead{} \\
   \raisebox{1.5ex}[-1.5ex]{Species} &
   \colhead{{ [km\,s$^{-1}$]} } &
   \colhead{{ [km\,s$^{-1}$]} } &
   \colhead{{ [cm$^{-2}$]} } &
   \raisebox{1.5ex}[-1.5ex]{Notes} 
}
 \startdata
\cutinhead{HS\,0033$+$4300, $v\approx-170$ \kms\ component}
  O\,I    &  \nodata   & \nodata   &$ <14.15       $&  upper limit \\
  Si\,II  &  \nodata   & \nodata   &$ <13.2        $&$\lambda 1190,1193,1526$; simulfit \\
  Si\,III &$ -142\pm3 $&$ 68\pm4  $&$ 13.5\pm0.3   $&  AOD, Voigt \\
  Si\,IV  &$ -177\pm2 $&$ 53\pm18 $&$ 13.5\pm0.1   $&$\lambda 1402$ only; AOD, Voigt \\
  C\,II   &$ -155:    $&$  65:    $&$  14.5:       $&  very uncertain, Voigt, AOD \\
  C\,IV   &$ -176\pm4 $&$ 35\pm7  $&$  14.13\pm0.07$&  doublet; AOD, Voigt \\
  N\,V    &  \nodata   & \nodata   &$ <13.56       $&  upper limit \\
\cutinhead{HS\,0033$+$4300, $v\approx-350$ \kms\ component}
  O\,I    &  \nodata   & \nodata   &$ <14.15       $&  upper limit \\
  Si\,II  &  \nodata   & \nodata   &$ <13.2        $&  upper limit \\
  Si\,III &$ -350\pm5 $&$ 22:     $&$  12.9:       $&$<3\sigma$ Voigt, AOD \\
  Si\,IV  &  \nodata   & \nodata   &  \nodata       &$\lambda 1402$; instrumental feature \\
  C\,II   &$ -359\pm11$&$ >50     $&$  14.3\pm0.1  $&  AOD, Voigt \\
  C\,IV   &$ -323\pm21$&$ 75\pm12 $&$  13.77\pm0.09$&  doublet; AOD, Voigt \\
  N\,V    &  \nodata   & \nodata   &$ <13.56       $&  upper limit \\
\cutinhead{HS\,0058$+$4213, $v\approx-180$ \kms\ component}
  O\,I    &  \nodata   &  \nodata  &$ <13.9        $&   upper limit \\
  Si\,II  &$ -188\pm11$&$ 37\pm15 $&$  13.8\pm0.2  $&$\lambda 1190,1193,1526$; AOD, Voigt \\
  Si\,III &$ -179\pm10$&$ 15\pm5  $&$  14.5\pm0.3  $&  Voigt \\
  Si\,IV  &$ -183\pm8 $&$ 38\pm7  $&$  13.45\pm0.18$&  doublet; AOD, Voigt \\
  C\,II   &$ -183\pm2 $&$ 33\pm6  $&$  14.5\pm0.1  $&  AOD, Voigt \\
  C\,IV   &$ -162\pm8 $&$ 53\pm14 $&$  14.15\pm0.13$&  doublet; AOD, Voigt, simulfit \\
  N\,V    &   \nodata  & \nodata   &$ <13.4        $&  upper limit
\enddata
 \tablenotetext{a}{All line profiles have been shifted by $+35$ \kms\ as discussed in the text.}
\end{deluxetable*}
\section{Results}
\label{sec: results}
\subsection{Q\,0045$+$3926}
\begin{figure}[htb!]
  \includegraphics[width=1\hsize]{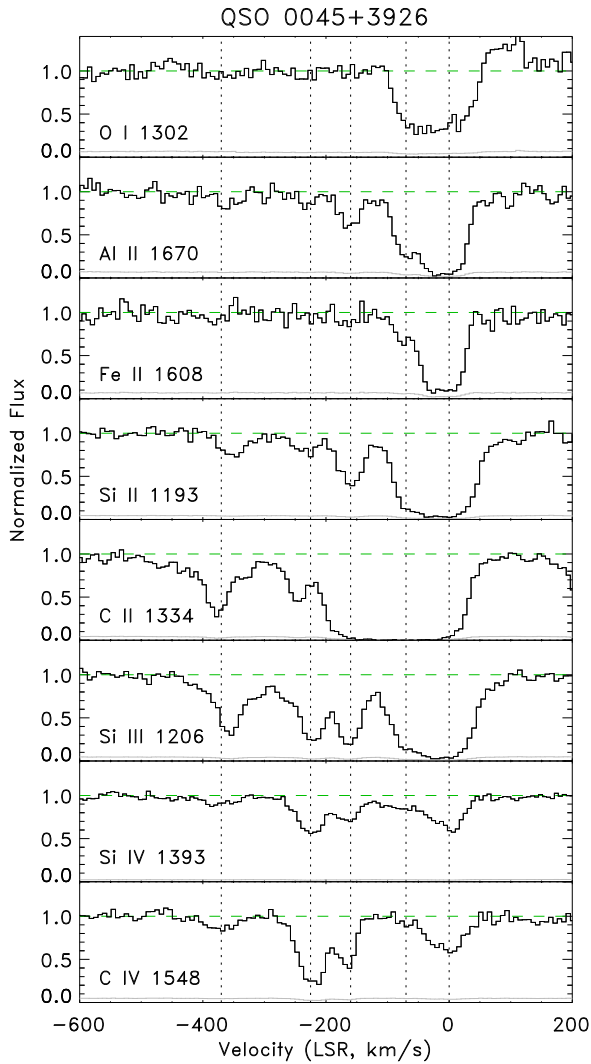}
  \caption{Normalized absorption profiles in various transitions along
  the Q\,0045$+$3926 sight line.  Data { are binned to three pixels ($\sim$40\% of the COS resolution) in all cases. The}
 (small) normalized error vector is shown in gray.  Vertical dashes
  correspond approximately to the velocity component centroids
  discussed in the text.} \label{fig:q0045_stack}
\end{figure}

The spectrum of Q\,0045$+$3926 shows 
 five distinct absorption components in the range $-400\la v_{lsr} \la 0$ \kms\ in many metal absorption lines from low-ionization \AlII\ and \FeII\ to highly-ionized \CIV\ transitions (Fig.~\ref{fig:q0045_stack}). 
The strong  absorption at $v\approx0$ km\,s$^{-1}$
   is similar to that seen in all other extragalactic UV absorption lines and can be unambiguously associated with absorption from the MW disk gas.  Other absorption, often blended with the MW gas, appears at 
  $v\approx-70$ \kms.  This component could plausibly be associated with the M\,31 halo given the overlap of MW and M\,31 stellar velocities discussed in Sect. 4.1 below.  
  However, this absorption is also consistent with infalling or outflowing gas within the Milky Way halo in the form of the high velocity clouds (HVCs) 
   and we deem this to be a more likely explanation (e.g., Wakker \& van Woerden 1997; Lehner et al. 2012).

At $v=-170$ and $-230$ \kms, two strong absorption components can be seen in transitions of \SiII\ (1193, 1190, 1526 \AA), \SiIII\ (1206.5 \AA), \SiIV\ (1393, 1402 \AA), and \CIV\ (1548, 1550 \AA).  The strong \SiII\ 1260 \AA\ line is blended with weaker \SII\ 1259 \AA\ 
absorption from the MW at that velocity, though the observed profile is consistent with the same absorption seen in three weaker transitions of the same ion.  The strong \CII\ 1334.5 \AA\ line is similarly complicated by HVC absorption from the excited-state \CII* 1335.7 \AA\ transition, but is again consistent with the absorption component structure seen at $v\sim-230$ \kms\ in other metal ions.  Notably, the relative strengths of the two lines in low ions as compared with higher ions in the same species (e.g., \SiII\ and \SiIV) show opposite trends implying that the material at $v\sim-170$ \kms\  is less ionized than the gas at $v\sim-230$ \kms.  

Prominent absorption is seen at $v\approx-370$ \kms\ 
in \SiII\ 1260 \AA, \CII, and \SiIII\ and weakly in other \SiII\ transitions as well as both \SiIV\ lines and \CIV\ 1548 \AA.  
{ This is a strong, distinct, system that is $\sim -70$ km\,s$^{-1}$ relative to the M\,31 systemic velocity, and 
we associate it with the GSS, despite the lack of obvious \HI\ emission in the immediate GSS, given the lower column densities that 
our observations sample.  
We will further discuss its possible origin and connections to stellar kinematics in its surroundings in Sect.~4.1.}

{ We note that all the detected three high velocity components could in principle arise in Milky Way HVCs as well, although we deem this very unlikely:}
MW halo gas clouds are typically observed at velocities up to $\sim200$ \kms, and the extreme wing of the distribution reaches up to 400 \kms\  \citep{WakkervanWoerden91,Putman02}.   
Some weak metal-line absorption systems have been detected in the MW halo that are not associated with 21-cm emission. These detections have come in both low-ionization species such 
as \CaII\ and \NaI\ \citep{Richter05,Richter09} and higher ions, such as \CIV\ and \OVI\ \citep{Sembach03}. 
\subsection{HS\,0033$+$4300 and HS\,0058$+$4213}
\begin{figure*}[t!]
  \includegraphics[width=0.5\hsize]{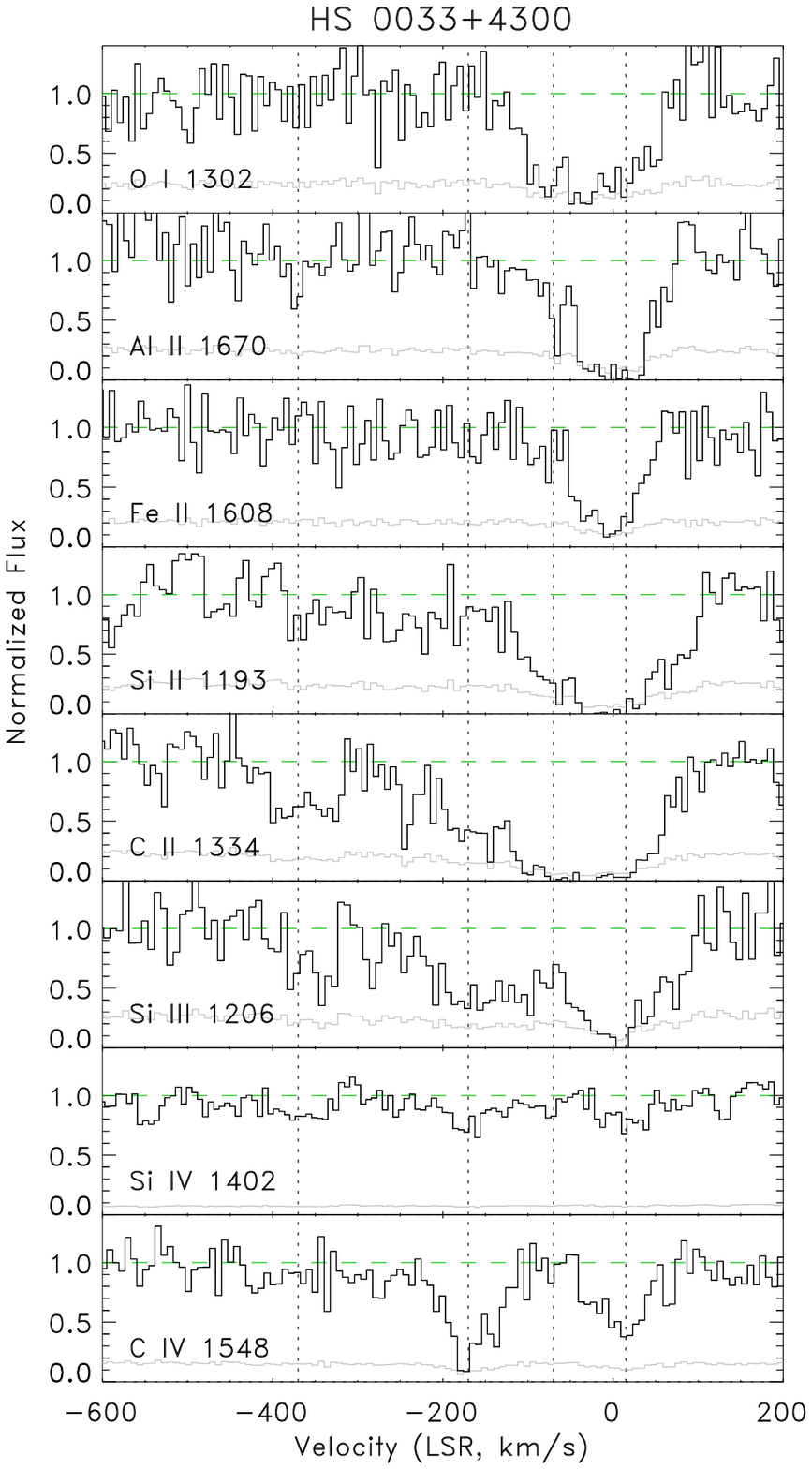}
  \includegraphics[width=0.5\hsize]{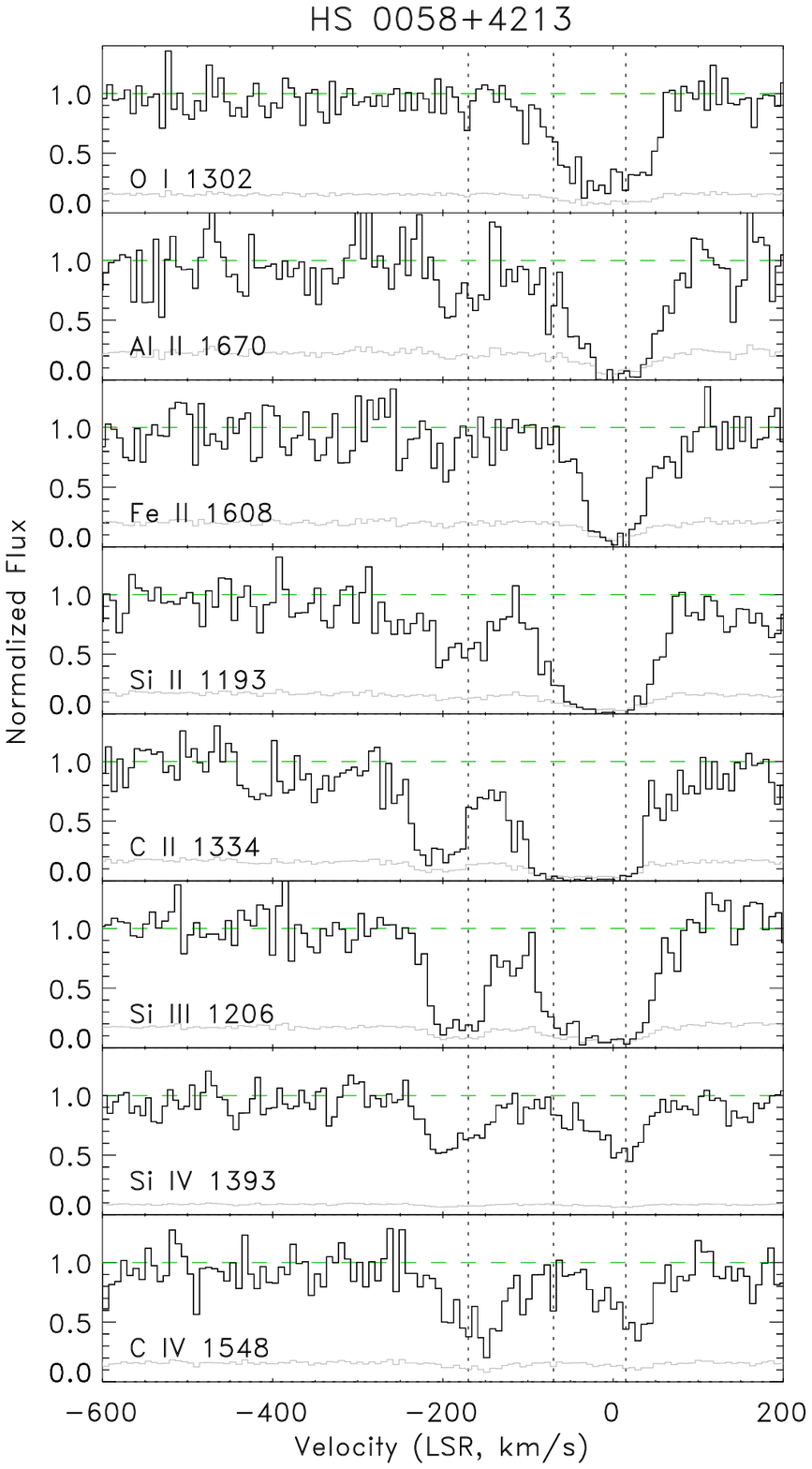}
  \caption{Same as Fig.~\ref{fig:q0045_stack} for HS\,0033$+$4300
  (left) and HS\,0058$+$4213 (right).  Data are binned by three pixels.
 The stronger Si\,IV line ($\lambda=1393$) is blended in the left
 panel with strong intrinsic absorption, so the weaker $\lambda=1402$
 line is shown. Vertical lines are the five components seen in Q\,0045+3926.} \label{fig:hs0033_stack}
\end{figure*}

The data for HS\,0033$+$4300 and HS\,0058$+$4213 are considerably noisier than those of Q\,0045+3926 due to fainter source fluxes and shorter exposure times.  
Detailed fits to the absorption profiles in these two sight lines are difficult, but similar trends are qualitatively seen.  Fig.~\ref{fig:hs0033_stack} shows absorption at $v\sim+15$ and $-70$ \kms\ in all bands which we ascribe to MW disk and halo gas. 
 Both sight lines show absorption at $v\sim-180$ to $-170$ \kms\ in the moderately and highly-ionized species as well, presumably tracing M\,31-velocity gas despite their considerable distance from the main body of the disk.  Both the MW and M\,31 absorption may plausibly be sub-divided into additional components as seen in the higher-quality data of Q\,0045+3926.

HS\,0033$+$4300 shows a low-significance absorption feature in several ions (\SiIII, \CIV) consistent with $v\sim-350$ \kms\ gas, though there is little evidence for this absorption in lower-ionization species.  This 
 contrasts Q\,0045+3926 where the strongest absorption at this velocity is seen in the low-ionization species \SiII, \SiIII, and \CII.  No convincing absorption is seen in the HS\,0058$+$4213 sight line at $v\la-300$ \kms.
\section{Discussion}
Our examination of the three absorption spectra finds that all have in common Galactic components, and components at 
around $-170$ to $-230$ km\,s$^{-1}$ that are plausibly associated with the M\,31 disk.  
\subsection{Gas Kinematics and Comparison with Stellar Samples}
Q 0045+3926 stands out in exhibiting a unique component at $-370$ km\,s$^{-1}$ not seen in the other two spectra, although there is a low-significance absorption at 
$-350$ km\,s$^{-1}$ 
toward HS\,0033$+$4300.  
The simulations  of Mori \& Rich (2008) suggest that { a stream component would be found in this NW region}, but the putative components appear too weak for a firm association.
As the sightline towards Q\,0045+3926 passes { $\sim$2.7 kpc} from the GSS we now turn to the question of whether observed stellar radial velocities or modeled stellar stream velocities are 
consistent with the $-370$ km\,s$^{-1}$ component in the gas.  
Fig.~4 (bottom panel) shows stellar radial velocities derived from Keck/DEIMOS measurements { within 7 kpc}  of Q\,0045+3926 (Ibata et al. 2015 in prep.); 
the stream shows a clear signature with a peak near $-500$ km\,s$^{-1}$, well separated from the gas component.  
  \begin{figure}[htb!]
\includegraphics[width=1\hsize]{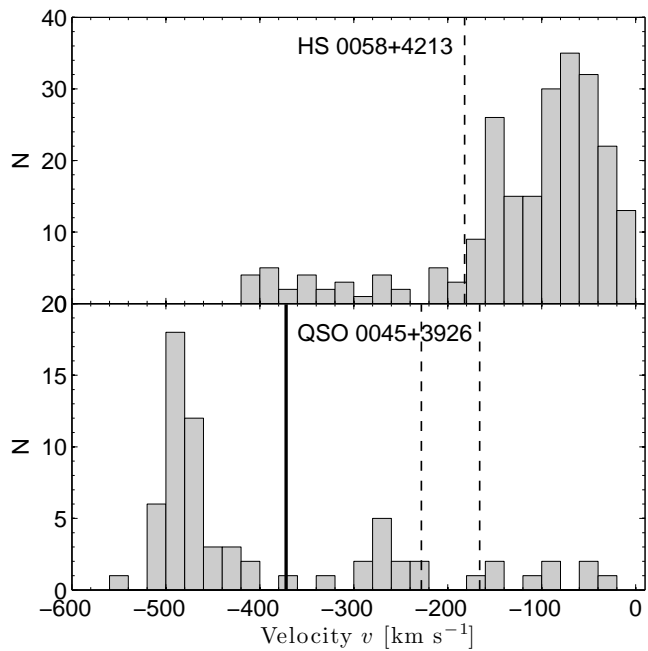}
  \caption{Observed 
  { stellar}
  radial velocity distributions in the immediate vicinity { (7 kpc)}
  of two of our { AGN sightlines,  where available}: HS 0058+4213 (top panel) and 
  Q 0045+3926 (bottom panel), taken from Ibata et al. (2004) and Ibata et al. (2015, in prep.). The bottom field contains a strong signature of the Giant Stream at $-500$ km\,s$^{-1}$. Solid and dashed lines indicate the detection of 
gas  absorption at those velocities, as listed in Tables~2,3.}
  \label{fig:starvel_obs}
\end{figure}
{ Such a difference is in line with the lack of any obvious spatial correlation between stellar and gas velocities reported by Lewis et al. (2013). The presence of a strong stellar velocity gradient within the GSS across its 4.5\degr~extent
has already been noted by Ibata et al. (2004), ranging from  $-300$ km\,s$^{-1}$ near its tip to $-500$ km\,s$^{-1}$ towards the disk of M31 (see also Koch et al. 2008; Gilbert et al. 2009). 
While the gas filaments also display a strong kinematic gradient, there are two marked differences (Lewis et al. 2013): firstly, gas is only detected with a significant parallel offset to the stellar stream and
secondly, gas near the disk shows the systemic velocity of M31, at   $-300$ km\,s$^{-1}$, transiting to more positive velocities as it approaches M\,33. 
Koch et al. (2008) and Gilbert et al. (2009) also identify a kinematic signature of the GSS at $-440$ km\,s$^{-1}$ and thus much closer to the gas velocity of our present work. 
The respective fields (``a3'') are, however, located at $\sim$30 kpc from Q 0045+3926, adding further complexity to the interpretation of velocity gradients and offsets between stream and stellar 
dynamics and spatial coincidences. 
}

{ The top panel of} Fig.~5, in turn,  shows the Mori \& Rich (2008) model distribution of velocities for the M\,31 halo  and stream particles within 5 kpc around Q\,0045+3926. 
The models had been  tailored to reproduce the morphology and kinematics of the GSS and distinguished particles from the disk/bulge, halo, and the accreted satellite.  
First, we note that no disk or bulge particles, ejected during the merger, are found in this region.
Halo stars in the simulation show, per construction, the broad distribution of stellar velocities that also include the three absorbers we observe in the gas phase and in the stellar samples.  
Turning to the satellite particles that make up the stream in the simulations, there is a clear coincidence of the broad distribution with the observed gas velocities at $-370$ \kms.
\begin{figure}[htb!]
\includegraphics[width=1\hsize]{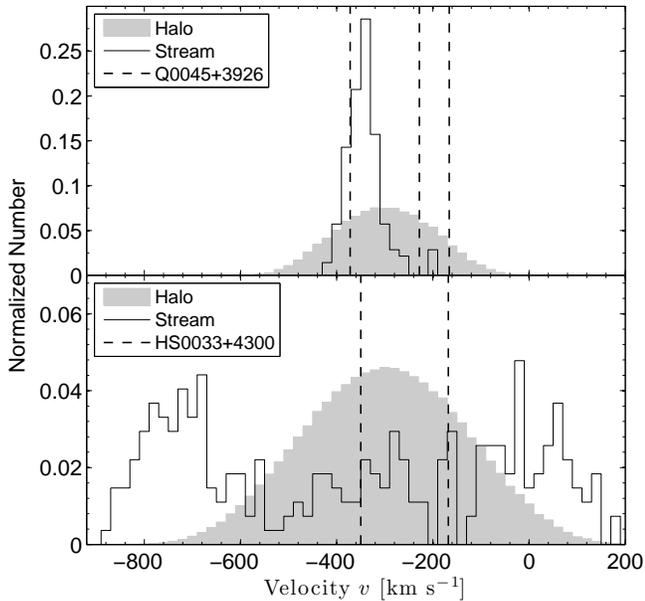}
  \caption{Same as Fig.~\ref{fig:starvel_obs}, but for the radial
  velocity distributions from the models of \citet{MoriRich08}.  
  { The gray-shaded and clear histograms are  drawn from the halo, and
  stream particles, respectively, while vertical lines indicate the detection of 
gas  absorption at those velocities in the respective sight lines, as listed in Table~2. 
No stream particles are present in the vicinity of HS 0058+4213.}
}\label{fig:starvel_sim}
\end{figure}
\addtocounter{footnote}{-1}
Moreover, the stream velocity distribution shows kurtosis\footnote{From the 140 stream particles within 5 kpc of Q\,0045+3926, we find a mean velocity of $-$340 km\,s$^{-1}$ 
with a 1$\sigma$-dispersion of 40 km\,s$^{-1}$, skewness of 1.2, and kurtosis of 3.1.}
 toward less negative velocities with a peak at $-300$ km\,s$^{-1}$. %

{
We now turn to the case of HS\,0058+4213.  
Fig.~4 (top) shows that the respective stellar sample was dominated by MW foreground dwarfs above $\sim -100$ km\,s$^{-1}$, which level off into the M31 
extended disk and halo profile below $\sim -150$ km\,s$^{-1}$; a tail of stars is found at  $-400$ km\,s$^{-1}<  v < -300$ km\,s$^{-1}$, which represents the canonical  halo population. 
The observed gas absorption in this sight line at $-180$ km\,s$^{-1}$ is also consistent with velocities of individual gas clouds in the M31 halo (e.g., Lewis et al. 2013) 
so that both stellar and gas velocities in this region probe M31's halo with no obvious connection to any substructures. We note that no Stream particles are found 
in this region of the Mori \& Rich (2008) simulations and the only dominant component is the broad, halo model peak at $-300$ km\,s$^{-1}$. 

Finally, while no stellar velocities are available in the vicinity of HS 0033+4300, the simulations predict that a rather diffuse stream component
 should be 
seen at highly negative velocities and another peak that blends into the MW foreground, above $\sim$0 km\,s$^{-1}$ (Fig.~5, bottom).
The gas absorption seen at $-170$ and, more weakly,  $-350$ km\,s$^{-1}$ thus appears to be consistent with typical halo gas.  
}

All three sight lines thus show absorption at near the systemic velocity of the M\,31 halo ($v\sim-170$ \kms), considering the stellar halo's broad velocity dispersion (at $\sim$85 km\,s$^{-1}$; Reitzel \& Guhathakurta 2002).  
Q\,0045$+$3926 shows two clear components separated by $\sim60$ \kms\ and it is clear from the ratios of high ions to low ions 
{ that these components have different ionization properties in different sightlines, 
in that the $v\sim -170$ km\,s$^{-1}$ system shows weaker C\,{\sc IV} and stronger low-ionization lines than the one $v\sim -230$ km\,s$^{-1}$.}
 Similarly, towards HS\,0033$+$4300, \CIV\ and \SiIV\ absorption is seen near $v\sim-176$ \kms\ while \CII\ and \SiIII\ show absorption centroids at $v\sim-150$ \kms.  

In conclusion, we note that stars are found with velocities consistent with all the three absorbing systems in the vicinity of the AGN sight lines, which is a manifestation of the large velocity dispersion of the M\,31 halo system. 
Based on the observed stellar velocities  and the comparison with the simulations of \citet{MoriRich08} we { argue }
that
the lowest-velocity component present in the AGN spectra are associated with the Stream itself.
\subsection{Modeling Physical Conditions}
With our  measurements of many metal ion column densities, it is possible to constrain the physical conditions in the absorbing gas with simple photoionization models.  To this end, we compare measured column densities with a grid of CLOUDY models developed for circumgalactic absorbers \citep[described fully in][]{Keeney13}.  Briefly, a slab of gas is illuminated by an ionizing continuum given by the extragalactic radiation field of \citet{HaardtMadau12}.  As in Keeney et al. (2013), we use a purely-extragalactic (AGN) background since all three of our sight lines probe regions outside M\,31's proximity zone where ionizing flux from early-type stars is expected to have an effect \citep[$D_{\rm prox}\la20$ kpc; ][]{GirouxShull97}.
  We calculate a grid of models where free parameters ionization parameter $U\equiv n_\gamma/n_H$ and metallicity $Z$ are varied to provide a grid of model conditions.  Best-fit model parameters for the three absorption components toward Q\,0045$+$3926 are listed in Table~4.       
\begin{deluxetable*}{ccccccccc}[htb!]
  \tabletypesize{\footnotesize}
  \tablecolumns{9} 
  \tablewidth{0pt} 
  \tablecaption{Derived Physical Quantities from CLOUDY Models}
   \tablehead{
     \colhead{} &
     \colhead{ $\log\,N_{\rm HI}$\tnma} &
    \colhead{ } &
     \colhead{ } &
     \colhead{ $\log\,n_{\rm H}$} &
     \colhead{$\log\,T$ } &
     \colhead{ } &
     \colhead{$D_{\rm cl}$ } &
     \colhead{$\log\,M_{\rm cl}$\tnmc} \\    
     \raisebox{1.5ex}[-1.5ex]{Absorber} &
     \colhead{[cm$^{-2}$]} &
     \raisebox{1.5ex}[-1.5ex]{$\log\,U$} &
     \raisebox{1.5ex}[-1.5ex]{$\log\,Z/Z_{\odot}$\tablenotemark{b}} &
     \colhead{[cm$^{-2}$}] &
     \colhead{[K]} &
     \raisebox{1.5ex}[-1.5ex]{$\log\,f_{\rm HI}$} &
     \colhead{[kpc], line of sight} &
     \colhead{[M$_\odot$]}
     }
  \startdata  
Q\,0045$+$3926, {$-370$ \kms} & $15.96\pm0.33$ & $-3.05_{-0.10}^{+0.18}$ & $-0.13_{-0.25}^{+0.34}$  & $-3.02$ & $4.04$ & $ -2.06$ & $0.35$ & $2.86$ \\
Q\,0045$+$3926, {$-230$ \kms} & $15.36\pm0.30$ & $-2.63_{-0.15}^{+0.26}$ & $+0.48_{-0.37}^{+0.44}$ & $-3.45$ & $3.77$ & $ -2.28$ & $0.40$ & $2.58$ \\
Q\,0045$+$3926, {$-170$ \kms} &$ 16.02\pm0.48$ & $-2.99_{-0.24}^{+0.23}$ & $+0.18_{-0.35}^{+0.80}$ & $-3.08$ & $3.83$ & $ -1.96$ & $0.38$ & $2.88$ \\          
    {HS\,0033$+$4300, $-350$ \kms} &$ 17.3\pm0.4$ & $-3.0_{-0.2}^{+0.3}$ & $+0.1_{-0.5}^{+0.3}$ & $-3.1$ & $4.2$ & $-2.3$ & $17.0$ & $7.8$           \\
    {HS\,0033$+$4300, $-170$ \kms} & $15.9\pm0.9$ & $-2.5_{-0.4}^{+0.7}$ & $+0.2_{-0.9}^{+1.2}$ & $-3.6$ & $4.2$ & $-2.7$ & $4.9$   & $5.7$ \\
    {HS\,0058$+$4213, $-180$ \kms} & $17.4\pm0.3$ & $-3.1\pm0.1$             &  $-0.7_{-0.2}^{+0.3}$  & $-2.9$ & $4.1$ & $-2.0$ & $8.6$  & $7.1$ 
\enddata    
 \tablenotetext{a}{$N_{\rm HI}$ is determined by assuming that Si\,II/III/IV all arise in a single photoionized phase.}
 \tablenotetext{b}{{ For this, CLOUDY  adopts the Solar metallicity of $\log\,Z_{\odot}$=0.012  from Asplund et al.  (2005).}}
 \tablenotetext{c}{$M_{\rm cl}$ is the cloud mass (in solar masses) assuming a spherical cloud with density $n_H$ and diameter $D_{cl}$.}
\end{deluxetable*}

Measurements in three adjacent ionization states of a metal provide a strong constraint on the ionization parameter.  In these data the strongest constraints are provided by Si\,II/III/IV, though C\,II/IV ratios 
{ are generally consistent.}

\HI\ absorption at the velocity of M\,31 is unrecoverably blended with Galactic absorption which makes the metallicity of the absorbing gas difficult to constrain directly.  We set an upper limit of $N_{\rm HI}<5\times10^{17}~\rm cm^{-2}$ since the Q\,0045$+$3926\ sight line lies outside the lowest \HI\ 21 cm column density contour of \citet{Thilker04}.  However, we can infer \NHI\ for each component by finding the range of total column density $N_H$ over which all of the silicon ion column densities in a given component can be explained simultaneously in a single, purely-photoionized phase of a given metallicity $Z$.  This gives values for $N_H$ and $Z$.  These quantities, along with the hydrogen neutral fraction ($f_{HI}$, uniquely given by the ionization parameter above) determines the neutral hydrogen column density \NHI\ all within fairly large uncertainties \citep{Stocke07}.  The equilibrium temperature, $T$, is a value returned by the best-fit $U,Z$ model.  A line-of-sight cloud size is then inferred in turn as $D_{\rm cl}=n_{\rm H}\,f_{HI}/N_{\rm HI}$.  Assuming a spherical cloud of constant density, we calculate a cloud mass $M_{\rm cl}$.  

We note that many assumptions have gone into these models and quantities farther down the chain of logic should be taken correspondingly with appropriate caution.
Modeled ionization parameters are probably reasonably accurate while inferred cloud mass may be inaccurate by orders of magnitude or more.

The CLOUDY model best fit parameters are $\log\,U\approx-3.0$ for both the GSS gas and the component at $-170$ \kms,  typical of conditions seen in the MW halo (e.g., Shull et al. 2011).  The gas at $-230$ \kms\ 
appears to be more highly ionized ($\log\,U\approx-2.6$) and thus typical of gas in ``highly-ionized'' HVCs, which is in agreement with the qualitative estimates of the respective line strengths seen in Fig.~2.  

The derived metallicities for the three absorption components along the Q\,0045$+$3926 sight line are all consistent with solar or slightly super-solar metallicity, though with large uncertainties.  
{ 
In particular, for the two M\,31-velocity components which are kinematically unrelated to the GSS and presumably trace gas in the halo of M\,31 (see also Lehner et al. 2015), 
these values are higher than the average stellar metallicities in the, generally more metal-poor, halo. 
}
 \citet{Koch08} found stellar metallicities of [Fe/H]$\approx-1.4$ dex in fields at 20--40 kpc from the galactic center.
Stars in the Giant Stream, however, are known to be more  metal-rich than the surrounding halo stars (Brown et al. 2006; Koch et al. 2008). 
{ 
For instance, Ibata et al. (2001) reported on an average [Fe/H] of approximately $-0.7$ dex, while 
Bernard et al. (2015) list a median metallicity of $-0.35$ dex based on their HST photometry.
Moreover, the metallicity distributions of Gilbert et al. (2009) peak at $-0.2$ dex and 
indicate a significant population of stars with roughly Solar metallicity in fields dominated by GSS material.
These findings support  our measurement of Solar metallicity in the stream gas, especially if one considers all the different, employed measurement techniques in the studies, 
from stellar calcium triplet spectroscopy, to color magnitude diagram fitting, and our modeling of the gas phase metallicity. 
}

Given the poor quality of the data for the HS\,0058$+$4213 and HS\,0033$+$4300 sight lines we hesitate to place much weight on their model results.  However, they are presented in 
{ Table~4}
 for completeness.  
\section{Summary}
We have investigated three sight lines in the M\,31 halo using HST/COS and AGN as background sources.   
All of the sight lines exhibit components that we associate with the Milky Way, and as well, a component 
at $-170$ to $-230$ km\,s$^{-1}$ that we assign to the disk of M\,31.  In the sight line of Q\,0045+3926, which is
projected { $\sim$2.7 kpc} from M\,31's Giant Stellar Stream, we find \SII~and \CII~absorption at 
$-370$ km\,s$^{-1}$, the most negative detected velocities in our sample.  
{ Here, we argue that the $-370$ km\,s$^{-1}$ component in Q\,0045+3946 is certainly  associated with structures in M\,31,  
most likely with the GSS. 

The bulk of {\em stellar} velocities in the GSS covers a broad range, reflecting a strong radial velocity gradient across its body. 
Stars in the immediate vicinity of the AGN sightline move at near $-500$ km\,s$^{-1}$, while other stream components have been 
found at velocities closer to our observed gas velocity (Ibata et al. 2004, 2015; Koch et al. 2008; Gilbert et al. 2009), albeit at larger 
projected distances. 
Aided by simulations of the merger that formed the GSS, Fardal et al. (2008)  suggested that many of its morphological features 
could best be reproduced with a rotating disk-progenitor. In that case, a detachment of the stellar from the gas velocities would occur naturally. 
It is obvious that different mechanisms were at play, contrasting  the tidal disruption of the stellar component of the stream progenitor
with  the gas removal by ram-pressure stripping in M\,31's hot circumgalactic corona (Mayer et al. 2006, Lewis et al. 2013; Lehner et al. 2015). 
Likewise, the remarkable lack of coherence between stellar structures and presently detected large-scale gas in the halo
emphasizes the complexity of the  GSS formation and dynamics (Lewis et al. 2013). 
}

 { 
Further evidence for a stream-gas connection comes from the putatively high metallicity we derived for the gas phase. 
 The halo of M\,31 has long been known to have stars with metallicity similar to that of 47 Tuc  (Mould \& Kristian 1983; Rich et al. 1996). 
 Even higher metallicities were found in larger surveys, allowing for 
a better characterization of the main halo polluters. In fact, the HST-based age-metallicity relation of Bernard et al. (2015) showed 
that halo fields rich in substructures experienced very rapid chemical enrichment so that, by a redshift of $z=1$, Solar metallicities were in place. 
This kind of evolutionary  history is typically found in early-type (dwarf) galaxies (e.g.,  Layden \& Sarajedini 2000; Bonifacio et al. 2004; Hendricks et al. 2014), allowing one to place further constraints on the 
progenitor's identikit.}
This emphasises the importance to increase the density of observational { sampling of absorbers within  the inner 30 kpc of the M\,31 halo}. This will be eased in the future by the increasing number of known AGN at suitable 
redshifts and magnitudes behind the M\,31--M\,33 system, as listed, e.g., by Huo et al. (2013). 
\begin{acknowledgements}
AK gratefully acknowledges the Deutsche Forschungsgemeinschaft for funding from Emmy-Noether grant Ko 4161/1. The authors would like to thank K.R. Sembach for designing the observing program and 
D.B. Reitzel for helpful discussions. We also thank the anonymous referee for a swift and constructive report. 
\end{acknowledgements}
\end{document}